\documentclass[conference]{IEEEtran}
\IEEEoverridecommandlockouts
\usepackage{cite}
\usepackage{amsmath,amssymb,amsfonts}
\usepackage{algorithmic}
\usepackage{graphicx}
\usepackage{textcomp}
\usepackage{xcolor}
\usepackage{comment}
\usepackage{siunitx}
\usepackage{booktabs}
\usepackage{comment}
\usepackage{soul}
\usepackage{placeins}
\usepackage{seqsplit}
\usepackage{makecell}
\usepackage{enumitem}
\usepackage{babel}
\usepackage{url}
\usepackage{multirow}
\usepackage{array,ragged2e}
\def\BibTeX{{\rm B\kern-.05em{\sc i\kern-.025em b}\kern-.08em
    T\kern-.1667em\lower.7ex\hbox{E}\kern-.125emX}}

\begin{document}

\title{PQC-LEO: An Evaluation Framework for Post-Quantum Cryptographic Algorithms}

\author{
\IEEEauthorblockN{Callum Turino}
\IEEEauthorblockA{
\textit{Edinburgh Napier University} \\
Edinburgh, UK \\
callum.turino@napier.ac.uk}
\and
\IEEEauthorblockN{William J. Buchanan}
\IEEEauthorblockA{
\textit{Edinburgh Napier University} \\
Edinburgh, UK \\
B.Buchanan@napier.ac.uk}
\and
\IEEEauthorblockN{Owen Lo}
\IEEEauthorblockA{
\textit{Edinburgh Napier University} \\
Edinburgh, UK \\
O.Lo@napier.ac.uk}
\and
\IEEEauthorblockN{Christoph Thümmler}
\IEEEauthorblockA{
\textit{6GHI} \\
Leipzig, Germany \\
christoph.thuemmler@6gha.com}
}

\maketitle

%--------------------------------------------------------------------------------------
% Abstract and Keywords
\begin{abstract}
Advances in quantum computing threaten digital communication security by undermining the foundations of current public-key cryptography through Shor's quantum algorithm. This has driven the development of Post-Quantum Cryptography (PQC), a new set of algorithms resistant to quantum attacks. While NIST has standardised several PQC schemes, challenges remain in their adoption. This paper introduces the PQC-LEO framework, a benchmarking suite designed to automate the evaluation of PQC computational and networking performance across x86 and ARM architectures. A proof-of-concept evaluation was conducted to demonstrate the framework's capabilities and highlight its application in supporting ongoing research on the adoption of PQC algorithms. The results show that there is a greater performance reduction in implementing PQC methods with higher security on ARM architectures than on the x86 architecture. 
\end{abstract}

% Keywords

%--------------------------------------------------------------------------------------
% Document Imports
%##########################################################################################
\section{Introduction}
The continued advancement of quantum computing poses a significant risk to global digital communication systems. Public-key cryptographic algorithms such as RSA and ECC will no longer be secure if a sufficiently powerful quantum computer were to be developed \cite{singh_future_2024}. This vulnerability arises from Peter Shor's quantum algorithm \cite{shor_algorithms_1994, shor_polynomial_time_1997}, which provides a mechanism for solving integer factorisation and discrete logarithm problems, used by current public-key schemes, within polynomial time. As a result, the confidentiality, integrity, and authenticity of communications secured through conventional public-key cryptography would be compromised. This threat extends from everyday applications such as secure web browsing to essential communications such as financial transactions and government transmissions \cite{rosch_grace_analysis_2022}.

To address the threat posed by quantum computing, Post-Quantum Cryptography (PQC) has been introduced as a new set of cryptographic algorithms resistant to both classical and quantum attacks. These schemes are based on mathematical problems known to be NP-hard, even to a quantum computer, a characteristic that traditional public-key schemes do not possess \cite{dam_survey_2023}. To support widespread adoption, standardisation bodies such as the National Institute of Standards and Technology (NIST) have led efforts in evaluating proposed PQC schemes \cite{joseph_transitioning_2022}. Following four rounds of evaluation, NIST have selected five PQC algorithms for standardisation \cite{dustin_moody_status_2022, alagic_status_2025}.

Despite the progress in PQC standardisation efforts, several challenges still remain for their widespread adoption. These include the practical deployment of PQC schemes in resource-constrained environments, such as IoT networks. Additionally, the reliance on lattice-based cryptography in standardised schemes may pose a risk if future vulnerabilities are discovered in this approach \cite{dustin_moody_status_2022, iqbal_survey_2023}. Furthermore, some PQC schemes can also strain broader system and communication architectures by exceeding design and protocol limitations, necessitating further research in real-world deployments \cite{saribas_performance_2022, bozhko_performance_2023}. Addressing these challenges requires a comprehensive evaluation framework that can provide a reliable and automated way of gathering algorithmic performance data of PQC schemes across varying device types and environments. In doing so, a more accurate understanding of the practical requirements for PQC adoption can be gained.

This paper introduces the Post-Quantum Cryptographic Library Evaluation Operator (PQC-LEO)\footnotemark, a benchmarking framework designed to assess the computational and networking performance of PQC schemes across various system architectures. It automates test environment setup, performance data collection, and result parsing. Additionally, the framework supports benchmarking PQC, Hybrid-PQC, and classical algorithms when integrated into TLS 1.3, enabling real-world network performance testing. It utilises PQC implementations natively supported within OpenSSL 3.5.0 and those included within the Liboqs and OQS-Provider libraries developed by the Open Quantum Safe Project (OQS) \cite{open_quantum_safe_2025}. Crucially, the framework supports TLS performance evaluation across both virtual and physical networks, addressing a gap often seen in similar tools. Finally, a proof-of-concept was performed on x86 and ARM systems to demonstrate the framework's functionality across multiple architectures.

\footnotetext{https://github.com/crt26/PQC-LEO}

%##########################################################################################
\section{Preliminaries}

%--------------------------------------------------------------------------------------
\subsection{The Quantum Threat}
Quantum computing introduces a significant threat to existing cryptographic systems through two key quantum algorithms: Shor's Algorithm and Grover's Algorithm \cite{tom_quantum_2023}. Shor's algorithm \cite{shor_algorithms_1994, shor_polynomial_time_1997} enables efficient solutions to integer factorisation and discrete logarithms, breaking the core assumptions of RSA and ECC public-key schemes \cite{bavdekar_post_2023}. Grover's algorithm \cite{grover_fast_1996} provides a quadratic speedup for unstructured search problems, impacting the security of symmetric encryption and hashing algorithms. While Grover's algorithm presents a manageable threat through adjustments in algorithm parameters and key sizes, Shor's algorithm compromises the core security assumptions of public-key cryptography. This has created an urgent need to develop and adopt new public-key encryption algorithms capable of resisting both classical and quantum attacks \cite{dustin_moody_status_2022, singh_future_2024}.

%--------------------------------------------------------------------------------------
\subsection{Post-Quantum Cryptography}
To combat the quantum threat posed by Shor's algorithm, the field of Post-Quantum Cryptography (PQC) was introduced. PQC algorithms are designed to replace current public-key schemes, with digital signature and Key-Encapsulation Mechanisms (KEMs) algorithms, resistant to both classical and quantum attacks \cite{dam_survey_2023}. These schemes rely on mathematical concepts that are known to be NP-hard, even for quantum computers. Common algorithm categories include lattice-based, code-based, hash-based, multivariate-based, and isogeny-based \cite{yalamuri_review_2022}. Each category possesses distinct advantages and disadvantages, with certain categories offering higher levels of performance and others stronger security guarantees. To determine the most effective solution for PQC and improve its adoption within digital communications infrastructures, extensive standardisation initiatives have been established to evaluate the performance, security, and adaptability of the proposed schemes \cite{joseph_transitioning_2022}.

%--------------------------------------------------------------------------------------
\subsection{Standardisation Efforts}
Among various global PQC standardisation efforts, the most prominent has been led by NIST through its Post-Quantum Cryptography Competition and its Additional Signatures competition \cite{chen_standardisation_2025}. Initiated in 2016, the main competition has progressed through several rounds to identify algorithms suitable for widespread adoption. Upon completion of the main competition, five algorithms were selected for standardisation, which have now been formalised as Federal Information Processing Standards (FIPS). These include ML-KEM (FIPS 203), ML-DSA (FIPS 204) and SLH-DSA (FIPS 205), with Falcon and HQC undergoing finalisation \cite{dustin_moody_status_2022, nist_FIPS_2024, alagic_status_2025}. In parallel with the main competition, NIST launched the Additional Signatures competition to explore other digital signature schemes not based on lattices. The first call for submissions resulted in 40 candidates, of which 14 were selected to advance to the second round of evaluation \cite{alagic_status_2024}.

%--------------------------------------------------------------------------------------
\subsection{Quantum Robust Methods}
Several quantum-robust methods have been considered during NIST's PQC evaluation. Tables \ref{tab:pqc_keysize} and \ref{tab:pqc_keysize2} present the currently standardised NIST PQC methods, and a summary of the considered schemes is provided in this section.

\subsubsection{Lattice-based}
Lattice-based methods include NTRU (Nth degree TRUncated polynomial ring), Learning With Errors (LWE), Ring LWE, and Learning with Rounding. These schemes exhibit favourable characteristics for digital signatures, key exchange, and encryption. Prominent methods include CRYSTALS-Kyber (ML-KEM) for key encapsulation, and CRYSTALS-Dilithium (ML-DSA) and Falcon (FN-DSA) for digital signatures. Lattice-based schemes generally demonstrate high performance for key generation, encapsulation and decapsulation, signing, and verification. Furthermore, the schemes utilise relatively compact public keys, private keys, and ciphertexts compared to other PQC solutions \cite{alagic_status_2025, wang_lattice-based_2023}. 

\subsubsection{Hash-based/Symmetric-based}
This category includes Merkle signatures, SPHINCS+ (SLH-DSA), and Picnic, which are primarily used for digital signatures. These schemes do not support public-key encryption and rely on generating a large number of private/public key pairs, each of which is only used once. In their basic form, the schemes require efficient tracking of key pair usage to prevent reuse \cite{bavdekar_post_2023, singh_future_2024}. However, SLH-DSA does support a stateless mechanism for managing private key usage. Hash-based schemes typically feature small private and public keys, but relatively long ciphertexts and signatures. Compared to lattice-based methods, they have significantly longer key generation, signature creation, and verification times \cite{lohachab_comprehensive_2020, li_hash-based_2022}.

\subsubsection{Code-based}
Code-based methods have been studied for several decades and include schemes such as McEliece, BIKE, and HQC. HQC is a key encapsulation mechanism based on Hamming Quasi-Cyclic and supports three security levels: HQC-128, HQC-192 and HQC-256 \cite{kumari_post_quantum_2022}. Although not as fast as lattice methods, code-based methods such as BIKE and HQC demonstrate relatively good performance for key generation, encapsulation, and decapsulation. Their key sizes are manageable, but the resulting ciphertexts are larger than those of equivalent lattice-based methods. McEliece in particular suffers from relatively slow key generation times \cite{alagic_status_2025, gharavi_post_quantum_2024}.

\subsubsection{Multivariate}
Multivariate quadratic methods encompass schemes such as oil-and-vinegar, 3WISE, DME-Sign, HPPC (Hidden Product of Polynomial Composition), MAYO, PROV (PRovable unbalanced Oil and Vinegar), QR-UOV, SNOVA, TUOV (Triangular Unbalanced Oil and Vinegar), UOV (Unbalanced Oil and Vinegar), and VOX \cite{alagic_status_2024}. These schemes are well-suited for digital signatures, offering efficient signing and verification operations along with smaller signature sizes \cite{dey_progress_2023}.

\subsubsection{Isogenies}
Isogeny-based schemes include those built on supersingular elliptic curve isogenies. Although promising, they currently remain relatively slow compared to other methods. The SIKE (Supersingular Isogeny Key Encapsulation) scheme, which employs the Supersingular Isogeny Diffie–Hellman (SIDH) protocol, was a finalist in the main NIST competition. However, the scheme was withdrawn when a key recovery attack was outlined in \cite{castryck2023efficient}. Other schemes include CGL, CSIDH, and SQIsign \cite{alagic_status_2024}.

\begin{table}[!t]
\caption{PQC Digital Signature Key Sizes (Bytes) \cite{lohmiller_survey_2025}}
\label{tab:pqc_keysize}
\centering
\normalsize
\renewcommand{\arraystretch}{0.9} % row height
\setlength{\tabcolsep}{5pt} % column padding
\resizebox{\columnwidth}{!}{%
\begin{tabular}{l l l l l}
\toprule
\textbf{Signature Algorithms} & \textbf{Public key size} & \textbf{Private key size} & \textbf{Signature size} & \textbf{Security level} \\
\midrule
ML-DSA-44 & 1,312 & 2,528 & 2,420 & 1 \\
ML-DSA-65 & 1,952 & 4,000 & 3,293 & 3 \\
ML-DSA-87 & 2,592 & 4,864 & 4,595 & 5 \\
FN-DSA 512 & 897 & 1,281 & 690 & 1 \\
FN-DSA 1024 & 1,793 & 2,305 & 1,330 & 5 \\
SLH-DSA-SHA2-128f & 32 & 64 & 17,088 & 1 \\
SLH-DSA-SHA2-192f & 48 & 96 & 35,664 & 3 \\
SLH-DSA-SHA2-256f & 64 & 128 & 49,856 & 5 \\ \midrule
RSA-2048 & 256 & 256 & 256 &  \\
ECC P256 & 64 & 32 & 256 &  \\ \bottomrule
\end{tabular}%
}
\end{table}

\begin{table}[!t]
\caption{PQC KEM Key Sizes (Bytes) \cite{lohmiller_survey_2025}}
\label{tab:pqc_keysize2}
\centering
\normalsize
\renewcommand{\arraystretch}{0.9} % row height
\setlength{\tabcolsep}{5pt} % column padding
\resizebox{\columnwidth}{!}{%
\begin{tabular}{l l l l l}
\toprule
\textbf{KEM Algorithms} & \textbf{Public key size} & \textbf{Private key size} &\textbf{ Ciphertext size} & \textbf{Security level} \\
\midrule
ML-KEM-512 & 800 & 1,632 & 768 & 1 \\
ML-KEM-768 & 1,184 & 2,400 & 1,088 & 3 \\
ML-KEM-1024 & 1,568 & 3,168 & 1,568 & 5 \\
HQC-128 & 2,249 & 2,289 & 4,497 & 1 \\
HQC-192 & 4,522 & 4,562 & 9,042 & 3 \\
HQC-256 & 7,245 & 7,285 & 14,485 & 5 \\ \midrule
P256 ECDH & 65 & 32 & 65 &  \\
P384 ECDH & 97 & 48 & 97 &  \\
P521 ECDH & 133 & 66 & 133 &  \\
\bottomrule
\end{tabular}%
}
\end{table}

%--------------------------------------------------------------------------------------
\subsection{Adoption Challenges}
Despite recent progress in standardisation efforts, the adoption of PQC schemes still faces several challenges. Their increased computational complexity and larger key and ciphertext sizes make them difficult to implement in resource-constrained environments  \cite{schoffel_secure_2022}. Due to the limited computational and energy capabilities of IoT devices, PQC adoption requires careful planning and extensive real-world testing. Networking performance is also a critical factor, as increased wireless transmissions significantly impact energy consumption \cite{saribas_performance_2022, bozhko_performance_2023}. As standardised algorithms are adopted, continued research into their practical deployment remains essential, given current limitations and ongoing scheme evaluations.
%##########################################################################################
\section{Related Work}
Several benchmarking frameworks have been proposed to assess PQC deployment challenges. One such example is the work by Rios \textit{et al.} \cite{rios_towards_2025}, who developed a benchmarking framework to evaluate PQC TLS performance. Through the use of OpenSSL 3.3.1 and OQS-Provider, the framework automates testing and supports simulated network congestion. However, it lacks support for non-localhost and Hybrid-PQC testing. Commey \textit{et al.} \cite{commey_performance_2025} developed a benchmarking framework to evaluate the CPU performance of PQC schemes on Apple Silicon, x86, and ARM systems. Using the Python wrapper for Liboqs and the Python cryptography library, they measured execution time, communication size, message size impact, and resource requirement trade-offs. While the study offers valuable cross-platform insights, the framework lacks scalability, omits memory metrics despite Liboqs support, and relies on outdated OQS dependencies.

Abbasi \textit{et al.} \cite{abbasi_practical_2025} benchmarked the computational, networking, and energy performance of select PQC and Hybrid-PQC schemes across multiple system architectures. Whilst the authors provide a comprehensive framework, they do not produce a public implementation of the evaluation, use depreciated versions of OQS libraries, and conduct no testing of networked devices. Hanna \textit{et al.} \cite{hanna_comprehensive_2025} present a testbed for evaluating PQC performance in IoT networks using BLE and Wi-Fi. They employ Raspberry Pi devices as a client and IoT gateway, with an x86 laptop as the server, testing various KEM and signature combinations with differing certification validation modes. The study offers useful insight into PQC performance in wireless environments, but its analysis is limited by outdated libraries and PQC implementations. Fitzgibbon \textit{et al.} \cite{fitzgibbon_constrained_2024} evaluated PQC performance across x86 and ARM architectures. They collected computational performance data using Liboqs and TLS handshake performance over physically networked ARM devices using OQS-OpenSSL. While the work offers non-localhost TLS testing, its analysis is limited by incomplete handshake timing results and the use of depreciated libraries.

Although the discussed PQC evaluation frameworks have addressed several gaps in research, they still present key limitations. This work aims to address these challenges and support the evaluation of both standardised and emerging PQC schemes.

%##########################################################################################
\section{PQC-LEO Framework Design}
To support ongoing research into the integration of PQC algorithms within modern and emerging communication systems, the PQC-LEO framework provides a flexible and automated benchmarking suite. This section outlines its design and functionality, detailing its core components, supported testing scenarios, and practical use cases.

%--------------------------------------------------------------------------------------
\subsection{Framework Overview and Use Case}
The PQC-LEO framework is a comprehensive and modular benchmarking suite designed to assess PQC algorithms. It evaluates both computational and networking performance across a range of system architectures and environments. It leverages PQC and Hybrid-PQC implementations present in OpenSSL 3.5.0 and those provided by the Liboqs and OQS-Provider libraries developed by the OQS project \cite{open_quantum_safe_2025}. This approach allows for the evaluation of not only the PQC schemes standardised by NIST but also facilitates the testing of PQC schemes that are still under review.

\begin{figure}[h]
    \centering
    \includegraphics[width=1\linewidth]{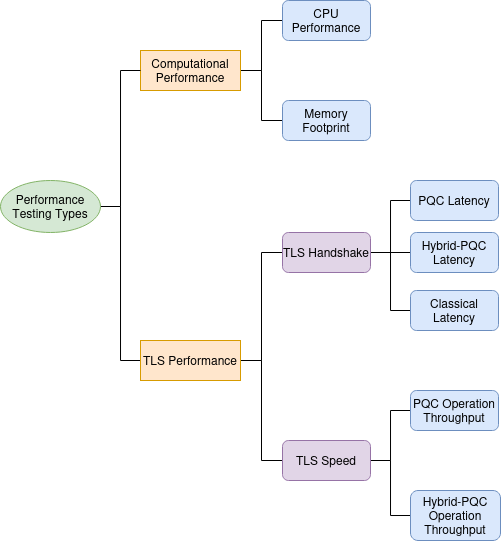}
    \caption{Overview of the PQC-LEO Framework Testing Types}
    \label{fig:framework-testing-overview}
\end{figure}

While the performance testing tools available in these libraries are widely used in PQC research, they require manual setup and configuration. This process can be time-consuming and difficult to scale, particularly when testing across varying system architectures and networking environments. Existing tools and proposed evaluation frameworks offer limited automation and testing across networked devices. The PQC-LEO framework addresses these limitations through full automation of environment setup, benchmarking, and result parsing. It supports testing on Debian-based systems using either an x86 or ARM CPU architecture and facilitates TLS performance testing across networked devices. An overview of the supported testing types can be seen in Figure \ref{fig:framework-testing-overview}.

%--------------------------------------------------------------------------------------
\subsection{PQC Computational Performance Testing}
The PQC-LEO framework provides functionality to benchmark the CPU and memory demands of various PQC algorithms supported by the Liboqs library. It evaluates both digital signature and KEM algorithms, collecting metrics such as CPU cycles, execution time, and memory usage. The process is streamlined by automating the execution of benchmarking routines for a user-defined number of test runs, ensuring consistency and repeatability. The resulting data is parsed into usable CSV formats using the framework's built-in tools, enabling straightforward analysis of PQC algorithmic performance on a given system.

%--------------------------------------------------------------------------------------
\subsection{PQC TLS Performance Testing}
The TLS testing functionality of the PQC-LEO framework enables benchmarking of algorithmic performance when implemented in the TLS 1.3 protocol. It supports both native implementations in OpenSSL 3.5.0 and those added through the OQS-Provider. The framework facilitates testing of PQC and Hybrid-PQC, alongside classical algorithms, such as RSA and ECC, to establish performance baselines. The TLS testing is divided into two sub-categories: TLS handshake performance and TLS cryptographic operation speed. TLS handshake performance tests measure connection establishment efficiency between a client and server machine, including scenarios with session ID reuse. The TLS speed tests assess the throughput and execution time of PQC algorithms when integrated into the OpenSSL library. Test coordination, execution, and data collection are fully automated for both test categories across a set of user-defined testing runs. Evaluations can be conducted locally via the localhost interface or across separate devices connected via a physical or virtual network, enabling scalable and realistic performance testing.

%--------------------------------------------------------------------------------------
\subsection{Known Limitations}
The PQC-LEO framework integrates a wide range of PQC algorithms through its dependencies, but some limitations arise from upstream constraints. HQC is disabled by default in OQS due to non-conformance with current specifications, though it can be re-enabled using the framework's advanced setup tools. On ARM systems, Falcon is supported for CPU and network performance testing, but memory testing is unavailable due to incompatibility with the Valgrind memory profiler. For TLS testing, several signature variants cannot be used in handshake evaluations due to TLS 1.3 non-conformance, though they remain available for speed tests. An exception to this is ML-DSA, which is supported for TLS handshake testing but is not currently functional for TLS speed tests. Additionally, while SLH-DSA is available within OpenSSL 3.5.0 for certificate generation, it is not yet functional for TLS testing; however, its non-standardised implementation, SPHINCS+, is still available for testing. These issues are expected to be resolved in future updates to the framework and its dependencies. Please refer to the supported algorithms documentation for full details \footnotemark.

\footnotetext{https://github.com/crt26/PQC-LEO/wiki/Algorithm-Support-Overview}
%##########################################################################################
\section{Implementation of PQC Computational Performance Testing}
The framework adopts a structured and automated approach to testing and data collection, enabling scalable evaluation of PQC computational performance. This section outlines the computational benchmarking implementation and the metrics it collects.

%--------------------------------------------------------------------------------------
\subsection{Computational Performance Evaluation Workflow}
The PQC computational performance benchmarking process is managed through a single bash script (\textit{pcq\_performance\_test.sh}), which orchestrates the CPU and memory profiling. Once executed, it handles environment configuration and prompts the user to specify testing parameters such as a machine identifier and the desired number of testing runs. This identifier isolates and labels collected performance results, facilitating comparisons across various systems. The performance benchmarking is divided into two stages: the CPU performance stage and the memory performance stage.

For CPU benchmarking, the framework invokes the \textit{speed\_kem} and \textit{speed\_sig} binaries included in the Liboqs library. These tools evaluate the computational efficiency of PQC KEM and digital signature algorithms by executing their respective cryptographic operations as many times as possible within a default three-second window. For memory benchmarking, the framework iterates over each algorithm and its operations, executing modified versions of the \textit{test\_kem\_mem} and \textit{test\_sig\_mem} binaries. These have been adjusted to suppress redundant output that could skew memory usage metrics. They are run under the Valgrind Massif Memory Profiler\footnotemark, which captures detailed memory allocation behaviour of individual cryptographic operations.

\footnotetext{https://github.com/fredericgermain/valgrind/tree/master}

%--------------------------------------------------------------------------------------
\subsection{Testing Output}
The collected performance results from both CPU and memory benchmarking are stored in a structured hierarchy under the \textit{test\_data/up\_results} directory. The results are organised by the assigned machine identifier and testing type. CPU performance results are saved in raw CSV format after each run. Memory profiling data from Massif is saved as text files containing snapshot data for the PQC algorithm's execution. These files are automatically converted into usable CSV formats using the Python parsing scripts included in the framework.

%##########################################################################################
\section{Implementation of PQC TLS Performance Testing}
The TLS implementation of the PQC-LEO framework provides a structured and automated approach for benchmarking PQC, Hybrid-PQC, and classical algorithms within TLS 1.3. This section outlines the execution process, including environment setup, coordinated testing between client and server machines, and the organisation and parsing of performance results.

%--------------------------------------------------------------------------------------
\subsection{TLS Certificate Generation and Test Controller}
As part of the framework's automated TLS testing functionality, the \textit{tls\_generate\_keys.sh} script streamlines the creation of all the required PQC, Hybrid-PQC, and classical X.509 certificates. This process prepares the testing environment and simplifies the collection of TLS handshake performance metrics by generating the certificates in advance. Following this, the \textit{pqc\_tls\_performance\_test.sh} script acts as the primary controller for orchestrating and executing TLS handshake and speed testing. It handles environment setup and prompts the user to input key testing parameters, including machine roles (server/client), number of test iterations, test-window durations, and machine IP addresses. A machine identifier is also assigned to tag and organise the collected performance data. Once configured, the script initiates TLS handshake tests on both client and server machines. If the system has been configured as a client, it will continue to execute the TLS speed tests once handshake testing concludes.

%--------------------------------------------------------------------------------------
\subsection{TLS Handshake Evaluation Workflow}
Once the environment is configured, the controller script invokes the appropriate internal handshake testing scripts based on the selected machine role. When set as the server, the framework utilises the OpenSSL \textit{s\_server} tool to host a TLS server using the specified cryptographic schemes. The client then uses the \textit{s\_time} tool to initiate as many TLS connections to the server as possible within a given time window, for both first use and reuse of a session ID. These tests are performed across multiple combinations of digital signature and KEM algorithms for PQC, Hybrid-PQC, and classical schemes, repeated over the configured number of iterations. A key component of the testing process is the automated control signalling mechanism implemented using the \textit{netcat} tool. It establishes a communication channel between the client and server machines for testing coordination. This mechanism manages test transitions, synchronises their start and end, prevents premature execution on the client side, and handles retries and errors. Crucially, it enables testing coordination between physically separate machines, allowing for fully unattended performance testing in real-world environments.

%--------------------------------------------------------------------------------------
\subsection{TLS Speed Evaluation Workflow}
After completing TLS handshake testing, the controller script calls the internal TLS speed testing tool. This tool evaluates PQC and Hybrid-PQC cryptographic operations within OpenSSL using the  \textit{speed} utility. The framework will conduct this testing for supported algorithms for the number of runs configured during the environment setup. This testing provides insight into the algorithm's standalone efficiency within OpenSSL, which may introduce additional overhead compared to the computational performance testing suite.

%--------------------------------------------------------------------------------------
\subsection{Testing Output}
The collected TLS handshake and speed performance results are stored in a structured hierarchy under the \textit{test\_data/up\_results} directory. The results are organised by the machine identifier assigned at the beginning of testing and then by test type. Handshake results are saved as plaintext logs and include metrics such as the number of connections per second in user-time and real-time, for both initial use and reuse of a session ID. The TLS speed results are also stored as plaintext files and contain metrics on the time to complete a cryptographic operation and the average throughput of operations per second. Once testing is completed, the framework automatically executes its parsing scripts to convert the raw performance outputs into usable CSV files.
%##########################################################################################
\section{Proof of Concept Evaluation}
To demonstrate the practical capabilities of the PQC-LEO framework, a proof-of-concept evaluation was conducted on both x86 and ARM system architectures. All performance evaluations used the framework's automated setup, testing, and parsing scripts, with no manual changes beyond the use of the supported advanced customisation flags.

%--------------------------------------------------------------------------------------
\subsection{Evaluation Environment}
To demonstrate the framework's ability to automate environment setup, each machine was prepared with a clean installation and minimal initial configuration. All tests in this evaluation were conducted using version 0.4.0 of the PQC-LEO framework, utilising all algorithms supported in that release. For tests requiring multiple devices, identical hardware models were used within each architecture to ensure consistency within the results. The full hardware and software specifications for each system type are outlined below:

\textbf{x86 Machine:} 
\begin{itemize}
    \item Machine Model - HP EliteDesk 800 G3 Mini 35w
    \item CPU - Intel i5-6500T, 4 cores @ 2.5GHz
    \item RAM - 8GB DDR4 @ 2133MT/s
    \item Operating System - Debian 12.11
    \item Compiler - gcc (12.2.0)
\end{itemize}

\textbf{ARM Machine:} 
\begin{itemize}
    \item Machine Model - Raspberry Pi 4 Model B
    \item CPU - Broadcom BCM2711 Cortex-A72 (ARM v8), 4 cores @ 1.8GHz
    \item RAM - 4GB LPDDR4 @ 3200MT/s
    \item Operating System - Raspberry Pi OS Lite (64-bit)
    \item Compiler - gcc (12.2.0)
\end{itemize}

To provide a comprehensive demonstration, the optional setup customisations built into the framework's setup script were used to re-enable HQC algorithms within the OQS libraries by supplying the \textit{--enable-all-hqc-algs} flag. For TLS performance testing, a Netgear GS308T switch was used to connect the devices. Additional system configuration and build information, including Liboqs build details for each system type, are provided in Tables \ref{tab:liboqs-build-x86} and \ref{tab:liboqs-build-arm}. This information is included because Liboqs is used directly in computational performance testing and provides the PQC implementations integrated into OpenSSL via the OQS-Provider library.

\begin{table}[ht]
\caption{Liboqs Build Configuration: x86 (Intel i5-6500T)}
\centering
\renewcommand{\arraystretch}{1.1}
\begin{tabular}{@{}lp{5.2cm}@{}}
\hline
OQS Version & 0.13.1-dev \\
Compiler & GCC 12.2.0 \\
Target & \texttt{x86\_64-Linux-6.1.0-37-amd64} \\
CMake & Release \\
Build Flags & \texttt{OQS\_DIST\_BUILD}, \texttt{OQS\_LIBJADE\_BUILD}, \newline \texttt{OQS\_OPT\_TARGET=generic} \\
Compile Options & \texttt{-O3}, \texttt{-fomit-frame-pointer}, \texttt{-Wa,--noexecstack}, \newline \texttt{-fdata-sections}, \texttt{-ffunction-sections}, \newline \texttt{-Wl,--gc-sections}, \texttt{-Wbad-function-cast} \\
OpenSSL & Enabled (v3.5.0, 8 Apr 2025) \\
Crypto Backends & AES: NI, SHA-2: OpenSSL, SHA-3: C \\
CPU Extensions & AVX2, AES, ADX, BMI1, BMI2, \newline PCLMULQDQ, POPCNT, SSE, SSE2, SSE3 \\
\hline
\end{tabular}
\label{tab:liboqs-build-x86}
\end{table}

\begin{table}[ht]
\caption{Liboqs Build Configuration: ARM (Raspberry Pi 4)}
\centering
\renewcommand{\arraystretch}{1.1}
\begin{tabular}{@{}lp{5.2cm}@{}}
\hline
OQS Version & 0.13.1-dev \\
Compiler & GCC 12.2.0 \\
Target & \texttt{aarch64-Linux-6.12.25+rpt-rpi-v8} \\
CMake & Release \\
Build Flags & \texttt{OQS\_SPEED\_USE\_ARM\_PMU}, \texttt{OQS\_DIST\_BUILD}, \newline \texttt{OQS\_LIBJADE\_BUILD}, \texttt{OQS\_OPT\_TARGET=generic} \\
Compile Options & \texttt{-march=armv8-a+crypto}, \texttt{-O3}, \texttt{-fomit-frame-pointer}, \newline \texttt{-Wa,--noexecstack}, \texttt{-ffunction-sections}, \newline \texttt{-fdata-sections}, \texttt{-Wl,--gc-sections}, \newline \texttt{-Wbad-function-cast} \\
OpenSSL & Enabled (v3.5.0, 8 Apr 2025) \\
Crypto Backends & AES: OpenSSL, SHA-2: OpenSSL, SHA-3: C \\
CPU Extensions & NEON \\
\hline
\end{tabular}
\label{tab:liboqs-build-arm}
\end{table}

%--------------------------------------------------------------------------------------
\subsection{Computational Performance Methodology}
The computational performance benchmarking evaluated the CPU and memory requirements of all supported PQC KEM and digital signature algorithms available in Liboqs. Each type of computational performance test was conducted over three runs, and the collected results are averaged across them. On both x86 and ARM systems, the Liboqs library was compiled to link to the OpenSSL 3.5.0 build generated by the framework's setup script. The x86 platform used the default liboqs flags alongside the \textit{no-shared linux\-x86\_64} flag. For ARM, the setup script enabled user-level access to the ARM Performance Monitoring Unit (PMU) via the PQAX\footnotemark library. Liboqs was then built using the \textit{-DOQS\_SPEED\_USE\_ARM\_PMU=ON} flag to instruct the testing binaries to use the ARM PMU.

\footnotetext{https://github.com/mupq/pqax}

%--------------------------------------------------------------------------------------
\subsection{TLS Performance Methodology}
To demonstrate the framework's ability to conduct real-world performance testing using physical network connections, two separate devices were used for each architecture category. The machines were connected via Ethernet and a Netgear GS308T switch. TLS handshake evaluations measured network latency when performing empty TLS handshakes for both initial use and reuse of a session ID, each within a 30-second testing window. TLS speed testing was also conducted over a 30-second testing window for both PQC and Hybrid-PQC algorithms. Both categories of testing were conducted over three iterations, with results averaged across the runs. Performance testing included all schemes supported natively in OpenSSL 3.5.0 and OQS-Provider, excluding those mentioned in the known limitations section. The OQS-Provider library was compiled using the same Liboqs build utilised during the computational performance testing. In addition to re-enabling HQC within OQS-Provider, all signature algorithms disabled by default in OQS-Provider were enabled through the setup script's configuration prompts. To accommodate this, the setup script dynamically adjusted hardcoded limits in the OpenSSL \textit{speed} tool's source code, as the default limit on registered algorithms is exceeded when enabling the full set of OQS-Provider signature schemes.
%##########################################################################################
\section{Collected Results}
For this proof-of-concept demonstration, the computational and TLS performance results were filtered to include only the top ten performing algorithms. The ranking criteria vary across testing sub-categories, with justifications provided in the following subsections. This selective approach enables a focused presentation of the framework's capabilities while maintaining a robust analytical methodology. Complete result sets are available in a dedicated GitHub repository created to display the results of this demonstration \footnotemark.

\footnotetext{https://github.com/crt26/PQC-LEO-PoC-Results}

%--------------------------------------------------------------------------------------
\subsection{Computational Performance Results}
In CPU testing, algorithms were ranked by calculating the lowest average operation time across the three operations. The top ten performing algorithms for the x86 and ARM systems can be seen in Tables \ref{tab:x86-cpu-performance} and \ref{tab:arm-cpu-performance}. These tables have been filtered to include only the total iterations, average operation time in microseconds, and average CPU cycles for each algorithm's operations. In memory testing, algorithms were ranked according to the lowest average peak memory footprint across the three operations. The x86 and ARM memory results are presented in Tables \ref{tab:x86-mem-performance} and \ref{tab:arm-mem-performance}, which have been filtered to display only the memory Heap, extHeap, and Stack values for each operation. For the purposes of this proof-of-concept, in instances where both standardised and non-standardised implementations of a given scheme were available, only the standardised version was included within the top-performing calculations. Additionally, on the ARM system, all instances of Falcon were excluded from memory performance rankings due to some of its variants not being supported by the Valgrind memory profiler.

%***************************************************************************************
% x86 CPU Performance
\begin{table*}[!t]
\caption{x86 PQC KEM and Digital Signature CPU Performance}
\label{tab:x86-cpu-performance}
\centering
\footnotesize
\renewcommand{\arraystretch}{0.6} % row height
\resizebox{\textwidth}{!}{%
\begin{tabular}{l lll lll lll}
\toprule
\multirow{2}{*}{\textbf{KEM Algorithms}} & \multicolumn{3}{c}{\textbf{keygen}} & \multicolumn{3}{c}{\textbf{encaps}} & \multicolumn{3}{c}{\textbf{decaps}} \\
 & \textbf{Iterations} & \textbf{Time (\si{\micro\second})} & \textbf{CPU Cycles} & \textbf{Iterations} & \textbf{Time (\si{\micro\second})} & \textbf{CPU Cycles} & \textbf{Iterations} & \textbf{Time (\si{\micro\second})} & \textbf{CPU Cycles} \\
\midrule
BIKE-L1 & 12,188 & 246 & 614,203 & 64,465 & 47 & 116,076 & 3,677 & 816 & 2,036,310 \\
BIKE-L3 & 4,298 & 698 & 1,742,118 & 28,205 & 106 & 265,365 & 1,162 & 2,583 & 6,446,158 \\
FrodoKEM-1344-AES & 2,009 & 1,494 & 3,728,852 & 1,556 & 1,929 & 4,813,579 & 1,597 & 1,879 & 4,688,521 \\
FrodoKEM-640-AES & 7,331 & 409 & 1,021,348 & 5,063 & 593 & 1,479,039 & 5,164 & 581 & 1,450,000 \\
FrodoKEM-640-SHAKE & 2,691 & 1,115 & 2,782,323 & 2,388 & 1,257 & 3,136,406 & 2,439 & 1,230 & 3,070,511 \\
FrodoKEM-976-AES & 3,503 & 857 & 2,138,216 & 2,719 & 1,104 & 2,754,646 & 2,838 & 1,057 & 2,638,509 \\
ML-KEM-1024 & 140,229 & 21 & 53,339 & 133,690 & 22 & 55,940 & 117,234 & 26 & 63,806 \\
ML-KEM-512 & 311,017 & 10 & 24,023 & 279,919 & 11 & 26,696 & 257,159 & 12 & 29,061 \\
ML-KEM-768 & 193,655 & 15 & 38,608 & 187,422 & 16 & 39,891 & 166,561 & 18 & 44,897 \\
sntrup761 & 10,243 & 293 & 731,032 & 177,213 & 17 & 42,187 & 161,373 & 19 & 46,339 \\
\midrule
\textbf{Digital Signature Algorithms} & \multicolumn{3}{c}{\textbf{keypair}} & \multicolumn{3}{c}{\textbf{sign}} & \multicolumn{3}{c}{\textbf{verify}} \\
\midrule
cross-rsdpg-128-fast & 238,478 & 13 & 31,346 & 8,619 & 348 & 868,617 & 14,019 & 214 & 534,087 \\
MAYO-1 & 36,962 & 81 & 202,483 & 15,121 & 198 & 495,093 & 31,908 & 94 & 234,583 \\
MAYO-2 & 54,898 & 55 & 136,294 & 27,415 & 109 & 273,039 & 85,545 & 35 & 87,427 \\
ML-DSA-44 & 95,712 & 31 & 78,169 & 36,149 & 83 & 207,149 & 101,431 & 30 & 73,766 \\
ML-DSA-65 & 55,419 & 54 & 135,033 & 22,502 & 133 & 332,677 & 58,616 & 51 & 127,682 \\
ML-DSA-87 & 35,319 & 85 & 211,944 & 18,099 & 166 & 413,641 & 37,091 & 81 & 201,806 \\
SNOVA\_24\_5\_4 & 21,532 & 142 & 353,636 & 6,358 & 479 & 1,195,722 & 25,028 & 123 & 307,955 \\
SNOVA\_24\_5\_4\_esk & 24,114 & 124 & 310,378 & 7,118 & 421 & 1,051,828 & 29,232 & 103 & 256,001 \\
SNOVA\_24\_5\_4\_SHAKE\_esk & 18,606 & 161 & 402,319 & 7,127 & 421 & 1,050,573 & 20,772 & 144 & 360,322 \\
SNOVA\_25\_8\_3 & 21,991 & 136 & 340,384 & 8,012 & 374 & 934,427 & 32,950 & 91 & 227,128 \\
\bottomrule
\end{tabular}%
}
\end{table*}

%***************************************************************************************
% ARM CPU Performance
\begin{table*}[!t]
\caption{ARM PQC KEM and Digital Signature CPU Performance}
\label{tab:arm-cpu-performance}
\centering
\footnotesize
\renewcommand{\arraystretch}{0.6} % row height
\resizebox{\textwidth}{!}{%
\begin{tabular}{l lll lll lll}
\toprule
\multirow{2}{*}{\textbf{KEM Algorithms}} & \multicolumn{3}{c}{\textbf{keygen}} & \multicolumn{3}{c}{\textbf{encaps}} & \multicolumn{3}{c}{\textbf{decaps}} \\
 & \textbf{Iterations} & \textbf{Time (\si{\micro\second})} & \textbf{CPU Cycles} & \textbf{Iterations} & \textbf{Time (\si{\micro\second})} & \textbf{CPU Cycles} & \textbf{Iterations} & \textbf{Time (\si{\micro\second})} & \textbf{CPU Cycles} \\
\midrule
BIKE-L1 & 71 & 42,850 & 8.66E+16 & 1,342 & 2,236 & 4,025,185 & 86 & 35,080 & 2.83E+11 \\
FrodoKEM-1344-SHAKE & 131 & 22,961 & 41,330,816 & 118 & 25,605 & 46,088,772 & 117 & 25,728 & 46,311,451 \\
FrodoKEM-640-AES & 177 & 16,988 & 30,577,889 & 174 & 17,282 & 31,108,065 & 175 & 17,207 & 30,972,405 \\
FrodoKEM-640-SHAKE & 533 & 5,633 & 10,139,587 & 470 & 6,392 & 11,505,317 & 473 & 6,353 & 11,434,769 \\
FrodoKEM-976-SHAKE & 237 & 12,644 & 22,759,808 & 214 & 14,082 & 25,347,324 & 215 & 13,985 & 25,172,295 \\
HQC-128 & 605 & 4,964 & 8,934,547 & 303 & 9,931 & 17,875,834 & 200 & 15,051 & 27,091,294 \\
ML-KEM-1024 & 31,758 & 94 & 169,889 & 29,548 & 102 & 182,609 & 26,963 & 111 & 200,169 \\
ML-KEM-512 & 72,443 & 41 & 74,437 & 64,588 & 46 & 83,500 & 61,870 & 48 & 87,190 \\
ML-KEM-768 & 45,702 & 66 & 118,046 & 42,982 & 70 & 125,527 & 39,627 & 76 & 136,167 \\
sntrup761 & 235 & 12,764 & 22,975,019 & 6,972 & 430 & 774,509 & 3,628 & 827 & 1,488,656 \\
\midrule
\textbf{Digital Signature Algorithms} & \multicolumn{3}{c}{\textbf{keypair}} & \multicolumn{3}{c}{\textbf{sign}} & \multicolumn{3}{c}{\textbf{verify}} \\
\midrule
cross-rsdp-128-fast & 52,543 & 57 & 102,689 & 1,776 & 1,689 & 3,040,862 & 3,064 & 979 & 1,762,731 \\
cross-rsdpg-128-balanced & 100,401 & 30 & 53,715 & 1,073 & 2,799 & 5,037,825 & 1,721 & 1,744 & 3,138,256 \\
cross-rsdpg-128-fast & 100,260 & 30 & 53,788 & 2,204 & 1,362 & 2,451,045 & 3,634 & 826 & 1,486,175 \\
MAYO-1 & 2,428 & 1,236 & 2,224,194 & 1,620 & 1,853 & 3,335,611 & 2,242 & 1,339 & 2,409,406 \\
MAYO-2 & 3,042 & 986 & 1,775,567 & 2,344 & 1,280 & 2,304,362 & 3,679 & 816 & 1,467,913 \\
ML-DSA-44 & 14,607 & 205 & 369,605 & 3,253 & 923 & 1,660,778 & 13,675 & 219 & 394,830 \\
ML-DSA-65 & 8,238 & 364 & 655,455 & 2,017 & 1,488 & 2,678,372 & 8,510 & 353 & 634,509 \\
ML-DSA-87 & 5,439 & 552 & 992,863 & 1,638 & 1,834 & 3,300,339 & 5,225 & 574 & 1,033,563 \\
SNOVA\_24\_5\_4\_esk & 3,485 & 861 & 1,549,715 & 1,109 & 2,706 & 4,870,013 & 2,634 & 1,139 & 2,050,716 \\
SNOVA\_25\_8\_3 & 2,922 & 1,027 & 1,848,193 & 1,258 & 2,385 & 4,293,079 & 3,432 & 874 & 1,573,458 \\
\bottomrule
\end{tabular}%
}
\end{table*}

%***************************************************************************************
% x86 Memory Performance
\begin{table*}[!t]
\caption{x86 PQC Memory Allocation at Peak Usage (Bytes)}
\label{tab:x86-mem-performance}
\centering
\footnotesize
\renewcommand{\arraystretch}{0.4} % row height
\setlength{\tabcolsep}{8pt} % column padding
\resizebox{\textwidth}{!}{%
\begin{tabular}{l lll lll lll}
\toprule
\multirow{2}{*}{\textbf{KEM Algorithms}} & \multicolumn{3}{c}{\textbf{keygen}} & \multicolumn{3}{c}{\textbf{encaps}} & \multicolumn{3}{c}{\textbf{decaps}} \\
 & \textbf{Heap} & \textbf{extHeap} & \textbf{Stack} & \textbf{Heap} & \textbf{extHeap} & \textbf{Stack} & \textbf{Heap} & \textbf{extHeap} & \textbf{Stack} \\
\midrule
BIKE-L1 & 7,884 & 52 & 92,056 & 9,489 & 79 & 26,456 & 9,521 & 87 & 77,944 \\
BIKE-L3 & 14,308 & 44 & 179,384 & 17,455 & 65 & 50,616 & 17,487 & 73 & 153,528 \\
FrodoKEM-640-AES & 30,816 & 40 & 43,208 & 40,552 & 64 & 84,168 & 40,568 & 72 & 104,776 \\
FrodoKEM-640-SHAKE & 31,488 & 40 & 37,960 & 41,224 & 64 & 58,752 & 41,240 & 72 & 79,328 \\
FrodoKEM-976-SHAKE & 48,912 & 40 & 55,808 & 64,680 & 64 & 89,664 & 64,704 & 80 & 121,280 \\
HQC-128 & 6,186 & 70 & 34,944 & 10,427 & 93 & 52,816 & 10,491 & 101 & 57,264 \\
HQC-192 & 10,740 & 60 & 66,912 & 19,526 & 82 & 102,960 & 19,590 & 90 & 111,968 \\
ML-KEM-1024 & 6,720 & 40 & 19,496 & 8,320 & 56 & 23,176 & 8,352 & 64 & 24,680 \\
ML-KEM-512 & 4,416 & 40 & 9,752 & 5,216 & 56 & 12,408 & 5,248 & 64 & 13,144 \\
ML-KEM-768 & 5,568 & 40 & 14,392 & 6,688 & 56 & 17,592 & 6,720 & 64 & 18,648 \\
\midrule
\textbf{Digital Signature Algorithms} & \multicolumn{3}{c}{\textbf{keypair}} & \multicolumn{3}{c}{\textbf{sign}} & \multicolumn{3}{c}{\textbf{verify}} \\
\midrule
Falcon-1024 & 9,778 & 94 & 35,912 & 11,340 & 132 & 83,024 & 7,028 & 116 & 10,248 \\
Falcon-512 & 7,858 & 94 & 19,032 & 8,710 & 122 & 42,064 & 8,710 & 122 & 2,072 \\
Falcon-padded-1024 & 9,778 & 94 & 35,912 & 11,158 & 122 & 82,992 & 6,846 & 106 & 10,248 \\
Falcon-padded-512 & 7,858 & 94 & 19,016 & 8,624 & 128 & 42,032 & 8,624 & 128 & 2,120 \\
ML-DSA-44 & 9,552 & 64 & 20,696 & 8,624 & 96 & 49,528 & 8,624 & 96 & 22,040 \\
SPHINCS+-SHAKE-128f-simple & 2,072 & 48 & 9,928 & 19,260 & 76 & 8,936 & 22,964 & 92 & 2,120 \\
SPHINCS+-SHAKE-128s-simple & 2,072 & 48 & 10,376 & 10,028 & 76 & 9,320 & 13,732 & 92 & 2,120 \\
SPHINCS+-SHAKE-192f-simple & 2,120 & 48 & 16,200 & 37,884 & 76 & 14,504 & 41,588 & 92 & 2,120 \\
SPHINCS+-SHAKE-192s-simple & 2,120 & 48 & 16,904 & 18,444 & 76 & 15,080 & 22,148 & 92 & 2,120 \\
SPHINCS+-SHAKE-256s-simple & 2,168 & 48 & 25,544 & 32,060 & 76 & 22,760 & 35,764 & 92 & 2,120 \\
\bottomrule
\end{tabular}%
}
\end{table*}

%***************************************************************************************
% ARM Memory Performance
\begin{table*}[!t]
\caption{ARM PQC Memory Allocation at Peak Usage (Bytes)}
\label{tab:arm-mem-performance}
\centering
\footnotesize
\renewcommand{\arraystretch}{0.4} % row height
\setlength{\tabcolsep}{8pt} % column padding
\resizebox{\textwidth}{!}{%
\begin{tabular}{l lll lll lll}
\toprule
\multirow{2}{*}{\textbf{KEM Algorithms}} &
\multicolumn{3}{c}{\textbf{keygen}} &
\multicolumn{3}{c}{\textbf{encaps}} &
\multicolumn{3}{c}{\textbf{decaps}} \\
& \textbf{Heap} & \textbf{extHeap} & \textbf{Stack} 
& \textbf{Heap} & \textbf{extHeap} & \textbf{Stack} 
& \textbf{Heap} & \textbf{extHeap} & \textbf{Stack} \\
\midrule
BIKE-L1 & 7,884 & 52 & 36,336 & 9,489 & 79 & 26,480 & 9,521 & 87 & 79,392 \\
BIKE-L3 & 14,308 & 44 & 69,664 & 17,455 & 65 & 50,944 & 17,487 & 73 & 156,816 \\
FrodoKEM-640-SHAKE & 31,488 & 40 & 37,072 & 41,224 & 64 & 58,800 & 41,240 & 72 & 79,408 \\
FrodoKEM-976-SHAKE & 48,912 & 40 & 55,904 & 64,680 & 64 & 89,728 & 64,704 & 80 & 121,216 \\
HQC-128 & 6,186 & 70 & 38,512 & 10,427 & 93 & 56,400 & 10,491 & 101 & 60,864 \\
HQC-192 & 10,740 & 60 & 70,944 & 19,526 & 82 & 107,008 & 19,590 & 90 & 116,032 \\
HQC-256 & 16,194 & 62 & 108,432 & 30,423 & 81 & 166,272 & 30,487 & 89 & 180,736 \\
ML-KEM-1024 & 6,720 & 40 & 19,040 & 8,320 & 56 & 22,720 & 8,352 & 64 & 24,240 \\
ML-KEM-512 & 4,416 & 40 & 9,280 & 5,216 & 56 & 11,968 & 5,248 & 64 & 12,688 \\
ML-KEM-768 & 5,568 & 40 & 13,968 & 6,688 & 56 & 17,136 & 6,720 & 64 & 18,176 \\
\midrule
\textbf{Digital Signature Algorithms} &
\multicolumn{3}{c}{\textbf{keypair}} &
\multicolumn{3}{c}{\textbf{sign}} &
\multicolumn{3}{c}{\textbf{verify}} \\
\midrule
cross-rsdp-128-fast & 5,789 & 67 & 9,072 & 24,321 & 95 & 116,128 & 20,009 & 79 & 57,360 \\
cross-rsdpg-128-fast & 5,766 & 74 & 4,816 & 17,846 & 106 & 77,712 & 13,534 & 90 & 42,512 \\
ML-DSA-44 & 9,552 & 64 & 38,704 & 8,016 & 96 & 53,360 & 8,016 & 96 & 36,912 \\
ML-DSA-65 & 11,664 & 64 & 60,672 & 11,017 & 87 & 81,040 & 11,017 & 87 & 58,464 \\
SPHINCS+-SHAKE-128f-simple & 0 & 0 & 7,776 & 22,964 & 92 & 1,584 & 22,964 & 92 & 1,584 \\
SPHINCS+-SHAKE-128s-simple & 0 & 0 & 7,776 & 13,732 & 92 & 1,744 & 13,732 & 92 & 1,744 \\
SPHINCS+-SHAKE-192f-simple & 0 & 0 & 7,776 & 41,588 & 92 & 1,584 & 41,588 & 92 & 1,584 \\
SPHINCS+-SHAKE-192s-simple & 0 & 0 & 7,776 & 22,148 & 92 & 1,584 & 22,148 & 92 & 1,584 \\
SPHINCS+-SHAKE-256f-simple & 1,560 & 48 & 8,576 & 55,828 & 92 & 1,744 & 55,828 & 92 & 1,744 \\
SPHINCS+-SHAKE-256s-simple & 1,560 & 48 & 8,832 & 31,452 & 76 & 6,560 & 35,764 & 92 & 1,584 \\
\bottomrule
\end{tabular}%
}
\end{table*}

%--------------------------------------------------------------------------------------
\subsection{TLS Handshake Results}
In the TLS handshake testing, the top ten performing algorithms were selected based on the highest number of TLS connections in real-time during the testing window. Real-time measurements were used over connections in user-time to provide a more realistic representation of algorithmic performance in real-world deployments. This is due to the metric capturing system-level delays and communication overheads between the two testing devices. TLS handshake performance results for the x86 machines are presented in Tables \ref{tab:x86-pqc-tls-handshake}, \ref{tab:x86-hybrid-tls-handshake}, and \ref{tab:x86-classic-tls-handshake}. Results for the ARM machines are shown in Tables \ref{tab:arm-pqc-tls-handshake}, \ref{tab:arm-hybrid-tls-handshake}, and \ref{tab:arm-classic-tls-handshake}. All tables have been filtered to display only the number of successful connections measured in real-time.

%***************************************************************************************
% x86 PQC TLS handshake
\begin{table}[]
\caption{x86 PQC TLS Handshake Performance}
\label{tab:x86-pqc-tls-handshake}
\centering
\small
\renewcommand{\arraystretch}{0.55} % same row height everywhere
\setlength{\tabcolsep}{2.5pt}      % same column padding everywhere
\resizebox{\columnwidth}{!}{%
\begin{tabular}{@{}l l r@{}}
\toprule
\textbf{Signing Algorithm} & \textbf{KEM Algorithm} & \textbf{TLS Connections} \\
\midrule
MLDSA44 & MLKEM512 & 7,111 \\
MLDSA65 & MLKEM512 & 6,313 \\
MLDSA87 & MLKEM512 & 8,041 \\
MLDSA87 & MLKEM768 & 6,334 \\
MLDSA87 & MLKEM1024 & 6,198 \\
CROSSrsdpg128balanced & MLKEM512 & 6,362 \\
CROSSrsdpg128fast & MLKEM512 & 6,802 \\
CROSSrsdpg192fast & MLKEM512 & 5,759 \\
OV\_Ip\_pkc\_skc & MLKEM512 & 6,129 \\
snova37172 & MLKEM512 & 7,429 \\
\bottomrule
\end{tabular}%
}
\end{table}

%***************************************************************************************
% x86 Hybrid-PQC TLS handshake
\begin{table}[]
\caption{x86 Hybrid-PQC TLS Handshake Performance}
\label{tab:x86-hybrid-tls-handshake}
\centering
\small
\renewcommand{\arraystretch}{0.55} % same row height everywhere
\setlength{\tabcolsep}{2.5pt}      % same column padding everywhere
\resizebox{\columnwidth}{!}{%
\begin{tabular}{@{}l l r@{}}
\toprule
\textbf{Signing Algorithm} & \textbf{KEM Algorithm} & \textbf{TLS Connections} \\
\midrule
p256\_mldsa44 & X25519MLKEM768 & 3,769 \\
p256\_mldsa44 & x25519\_bikel1 & 3,468 \\
p384\_mldsa65 & X25519MLKEM768 & 3,438 \\
p256\_mayo2 & X25519MLKEM768 & 3,745 \\
p256\_OV\_Ip\_pkc & X25519MLKEM768 & 5,464 \\
p256\_OV\_Ip\_pkc\_skc & X25519MLKEM768 & 4,028 \\
p256\_OV\_Ip\_pkc\_skc & SecP256r1MLKEM768 & 3,368 \\
p256\_snova37172 & X25519MLKEM768 & 5,176 \\
p256\_snova37172 & SecP256r1MLKEM768 & 3,647 \\
p256\_snova37172 & x25519\_mlkem512 & 3,344 \\
\bottomrule
\end{tabular}%
}
\end{table}

%***************************************************************************************
% x86 Classic TLS handshake
\begin{table}[]
\caption{x86 Classical TLS Handshake Performance}
\label{tab:x86-classic-tls-handshake}
\centering
\small
\renewcommand{\arraystretch}{0.55} % same row height everywhere
\setlength{\tabcolsep}{2.5pt}      % same column padding everywhere
\resizebox{\columnwidth}{!}{%
\begin{tabular}{@{}p{2cm} l r@{}}
\toprule
\textbf{Signing Algorithm} & \textbf{Ciphersuite} & \textbf{TLS Connections} \\ \midrule
RSA\_2048 & TLS\_AES\_256\_GCM\_SHA384 & 4,186 \\
RSA\_3072 & TLS\_AES\_256\_GCM\_SHA384 & 3,404 \\
prime256v1 & TLS\_AES\_256\_GCM\_SHA384 & 5,456 \\
RSA\_2048 & TLS\_CHACHA20\_POLY1305\_SHA256 & 3,985 \\
RSA\_3072 & TLS\_CHACHA20\_POLY1305\_SHA256 & 3,638 \\
RSA\_4096 & TLS\_CHACHA20\_POLY1305\_SHA256 & 2,871 \\
prime256v1 & TLS\_CHACHA20\_POLY1305\_SHA256 & 5,472 \\
RSA\_2048 & TLS\_AES\_128\_GCM\_SHA256 & 4,052 \\
RSA\_3072 & TLS\_AES\_128\_GCM\_SHA256 & 3,493 \\
prime256v1 & TLS\_AES\_128\_GCM\_SHA256 & 5,683 \\
\bottomrule
\end{tabular}%
}
\end{table}

%***************************************************************************************
% ARM PQC TLS handshake
\begin{table}[]
\caption{ARM PQC TLS Handshake Performance}
\label{tab:arm-pqc-tls-handshake}
\centering
\small
\renewcommand{\arraystretch}{0.6} % row height
\resizebox{\columnwidth}{!}{%
\begin{tabular}{@{}l l r@{}}
\toprule
\textbf{Signing Algorithm} & \textbf{KEM Algorithm} & \textbf{TLS Connections} \\ \midrule
MLDSA44 & MLKEM512 & 5,530 \\
MLDSA44 & MLKEM768 & 4,931 \\
MLDSA44 & MLKEM1024 & 4,795 \\
MLDSA65 & MLKEM512 & 4,130 \\
MLDSA65 & MLKEM768 & 3,712 \\
MLDSA65 & MLKEM1024 & 3,625 \\
MLDSA87 & MLKEM512 & 4,062 \\
MLDSA87 & MLKEM768 & 3,701 \\
MLDSA87 & MLKEM1024 & 3,672 \\
CROSSrsdpg128fast & MLKEM512 & 2,892 \\
\bottomrule
\end{tabular}%
}
\end{table}

%***************************************************************************************
% ARM Hybrid-PQC TLS handshake
\begin{table}[]
\caption{ARM Hybrid-PQC TLS Handshake Performance}
\label{tab:arm-hybrid-tls-handshake}
\centering
\small
\renewcommand{\arraystretch}{0.55} % same row height everywhere
\setlength{\tabcolsep}{2.5pt}      % same column padding everywhere
\resizebox{\columnwidth}{!}{%
\begin{tabular}{@{}l l r@{}}
\toprule
\textbf{Signing Algorithm} & \textbf{KEM Algorithm} & \textbf{TLS Connections} \\ \midrule
p256\_mldsa44 & X25519MLKEM768 & 2,778 \\
p256\_mldsa44 & SecP256r1MLKEM768 & 2,211 \\
p256\_falcon512 & X25519MLKEM768 & 2,867 \\
p256\_falcon512 & SecP256r1MLKEM768 & 2,236 \\
p256\_falconpadded512 & X25519MLKEM768 & 2,794 \\
p256\_falconpadded512 & SecP256r1MLKEM768 & 2,230 \\
p256\_mayo1 & X25519MLKEM768 & 2,233 \\
p256\_mayo2 & X25519MLKEM768 & 2,531 \\
p256\_OV\_Ip\_pkc & X25519MLKEM768 & 2,399 \\
p256\_snova2454esk & X25519MLKEM768 & 2,132 \\
\bottomrule
\end{tabular}%
}
\end{table}

%***************************************************************************************
% ARM Classic TLS handshake
\begin{table}[]
\caption{ARM Classical TLS Handshake Performance}
\label{tab:arm-classic-tls-handshake}
\centering
\small
\renewcommand{\arraystretch}{0.55} % same row height everywhere
\setlength{\tabcolsep}{2.5pt}      % same column padding everywhere
\resizebox{\columnwidth}{!}{%
\begin{tabular}{@{}p{2cm} l r@{}}
\toprule
\textbf{Signing Algorithm} & \textbf{Ciphersuite} & \textbf{TLS Connections} \\ \midrule
RSA\_2048 & TLS\_AES\_256\_GCM\_SHA384 & 1,855 \\
RSA\_3072 & TLS\_AES\_256\_GCM\_SHA384 & 1,000 \\
prime256v1 & TLS\_AES\_256\_GCM\_SHA384 & 4,937 \\
secp384r1 & TLS\_AES\_256\_GCM\_SHA384 & 1,257 \\
RSA\_2048 & TLS\_CHACHA20\_POLY1305\_SHA256 & 1,841 \\
prime256v1 & TLS\_CHACHA20\_POLY1305\_SHA256 & 4,992 \\
secp384r1 & TLS\_CHACHA20\_POLY1305\_SHA256 & 1,258 \\
RSA\_2048 & TLS\_AES\_128\_GCM\_SHA256 & 1,818 \\
prime256v1 & TLS\_AES\_128\_GCM\_SHA256 & 5,047 \\
secp384r1 & TLS\_AES\_128\_GCM\_SHA256 & 1,258 \\
\bottomrule
\end{tabular}%
}
\end{table}

%--------------------------------------------------------------------------------------
\subsection{TLS Speed Results}
TLS speed testing was filtered to report cryptographic operation throughput and was chosen over individual operation timing metrics to better reflect cryptographic performance under sustained loads. The top ten performing algorithms were selected by calculating the highest average performance across the three cryptographic operations. TLS speed results for the x86 system are presented in Tables \ref{tab:x86-pqc-tls-speed} and \ref{tab:x86-hybrid-tls-speed}, while results for the ARM system are shown in Tables \ref{tab:arm-pqc-tls-speed} and \ref{tab:arm-hybrid-tls-speed}.

%***************************************************************************************
% Row 1: x86 results side by side
\begin{table*}[t]
\centering
\begin{minipage}{0.49\textwidth}
\centering
\normalsize
\renewcommand{\arraystretch}{0.85}
\setlength{\tabcolsep}{7pt}
\caption{x86 PQC TLS Speed Performance}
\label{tab:x86-pqc-tls-speed}
\resizebox{\linewidth}{!}{%
\begin{tabular}{@{}p{45mm}rrr@{}}
\toprule
\textbf{KEM Algorithm} & \textbf{Keygen/s} & \textbf{Encaps/s} & \textbf{Decaps/s} \\
\midrule
frodo640aes & 2,375 & 1,644 & 1,715 \\
frodo640shake & 910 & 808 & 823 \\
frodo976aes & 1,172 & 879 & 912 \\
frodo1344aes & 671 & 502 & 520 \\
bikel1 & 4,046 & 21,710 & 1,224 \\
bikel3 & 1,425 & 9,406 & 386 \\
bikel5 & 549 & 4,440 & 161 \\
ML-KEM-512 & 28,365 & 45,502 & 30,000 \\
ML-KEM-768 & 18,441 & 34,804 & 22,510 \\
ML-KEM-1024 & 11,987 & 26,062 & 17,007 \\ 
\midrule
\textbf{Digital Signature Algorithm} & \textbf{Keygen/s} & \textbf{Sign/s} & \textbf{Verify/s} \\
\midrule
mayo2 & 17,736 & 9,161 & 28,479 \\
CROSSrsdp128balanced & 47,067 & 1,134 & 1,689 \\
CROSSrsdp128fast & 39,899 & 1,987 & 3,277 \\
CROSSrsdpg128balanced & 73,830 & 1,494 & 2,222 \\
CROSSrsdpg128fast & 77,192 & 2,894 & 4,540 \\
CROSSrsdpg128small & 73,626 & 733 & 1,164 \\
CROSSrsdpg192balanced & 43,241 & 1,016 & 1,495 \\
CROSSrsdpg192fast & 43,009 & 1,311 & 2,032 \\
OV\_Is & 669 & 20,748 & 38,832 \\
OV\_Ip & 1,105 & 25,711 & 31,286 \\ 
\bottomrule
\end{tabular}}
\end{minipage}\hfill
\begin{minipage}{0.49\textwidth}
\centering
\normalsize
\renewcommand{\arraystretch}{0.85}
\setlength{\tabcolsep}{7pt}
\caption{x86 Hybrid-PQC TLS Speed Performance}
\label{tab:x86-hybrid-tls-speed}
\resizebox{\linewidth}{!}{%
\begin{tabular}{@{}p{45mm}rrr@{}}
\toprule
\textbf{KEM Algorithm} & \textbf{Keygen/s} & \textbf{Encaps/s} & \textbf{Decaps/s} \\
\midrule
x25519\_frodo640aes & 2,159 & 1,439 & 1,504 \\
p256\_mlkem512 & 92 & 9,262 & 5,259 \\
x25519\_mlkem512 & 18,128 & 10,547 & 10,620 \\
x448\_mlkem768 & 3,788 & 2,160 & 2,157 \\
p256\_bikel1 & 90 & 7,033 & 1,000 \\
x25519\_bikel1 & 3,439 & 7,719 & 1,111 \\
x448\_bikel3 & 1,055 & 1,810 & 329 \\
X25519MLKEM768 & 10,253 & 8,990 & 11,605 \\
X448MLKEM1024 & 3,046 & 2,070 & 3,826 \\
SecP256r1MLKEM768 & 12,398 & 8,416 & 8,688 \\
\midrule
\textbf{Digital Signature Algorithm} & \textbf{Keygen/s} & \textbf{Sign/s} & \textbf{Verify/s} \\
\midrule
p256\_mldsa44 & 92 & 8,836 & 9,196 \\
rsa3072\_mldsa44 & 5 & 480 & 13,600 \\
p256\_falcon512 & 53 & 3,169 & 7,652 \\
rsa3072\_falcon512 & 5 & 443 & 10,489 \\
p256\_falconpadded512 & 53 & 3,145 & 7,630 \\
p256\_mayo2 & 91 & 7,078 & 8,711 \\
p256\_OV\_Is & 80 & 11,451 & 8,282 \\
p256\_OV\_Ip & 85 & 12,865 & 8,375 \\
p256\_OV\_Is\_pkc & 82 & 10,867 & 5,219 \\
p256\_OV\_Ip\_pkc & 86 & 12,125 & 6,041 \\
\bottomrule
\end{tabular}}
\end{minipage}
\end{table*}

%***************************************************************************************
% Row 2: ARM results side by side
\begin{table*}[t]
\centering
\begin{minipage}{0.49\textwidth}
\centering
\normalsize
\renewcommand{\arraystretch}{0.85}
\setlength{\tabcolsep}{7pt}
\caption{ARM PQC TLS Speed Performance}
\label{tab:arm-pqc-tls-speed}
\resizebox{\linewidth}{!}{%
\begin{tabular}{@{}p{45mm}rrr@{}}
\toprule
\textbf{KEM Algorithm} & \textbf{Keygen/s} & \textbf{Encaps/s} & \textbf{Decaps/s} \\
\midrule
frodo640aes & 59 & 58 & 58 \\
frodo640shake & 176 & 158 & 159 \\
frodo976shake & 79 & 71 & 72 \\
frodo1344shake & 44 & 39 & 39 \\
bikel1 & 24 & 447 & 28 \\
bikel3 & 8 & 146 & 9 \\
hqc128 & 201 & 101 & 67 \\
ML-KEM-512 & 12,529 & 19,796 & 13,076 \\
ML-KEM-768 & 8,285 & 14,865 & 9,516 \\
ML-KEM-1024 & 5,678 & 11,417 & 7,304 \\
\midrule
\textbf{Digital Signature Algorithm} & \textbf{Keygen/s} & \textbf{Sign/s} & \textbf{Verify/s} \\
\midrule
CROSSrsdp128balanced & 15,834 & 319 & 535 \\
CROSSrsdp128fast & 15,806 & 598 & 1,031 \\
CROSSrsdp128small & 15,791 & 158 & 259 \\
CROSSrsdpg128balanced & 27,396 & 361 & 588 \\
CROSSrsdpg128fast & 27,412 & 742 & 1,220 \\
CROSSrsdpg128small & 27,566 & 182 & 294 \\
CROSSrsdpg192balanced & 16,426 & 244 & 396 \\
CROSSrsdpg192fast & 16,365 & 330 & 521 \\
CROSSrsdpg192small & 16,283 & 130 & 205 \\
OV\_Ip & 172 & 6,421 & 8,165 \\
\bottomrule
\end{tabular}}
\end{minipage}\hfill
\begin{minipage}{0.49\textwidth}
\centering
\normalsize
\renewcommand{\arraystretch}{0.85}
\setlength{\tabcolsep}{7pt}
\caption{ARM Hybrid-PQC TLS Speed Performance}
\label{tab:arm-hybrid-tls-speed}
\resizebox{\linewidth}{!}{%
\begin{tabular}{@{}p{45mm}rrr@{}}
\toprule
\textbf{KEM Algorithm} & \textbf{Keygen/s} & \textbf{Encaps/s} & \textbf{Decaps/s} \\
\midrule
x25519\_frodo640shake & 175 & 150 & 151 \\
p256\_mlkem512 & 33 & 2,531 & 1,588 \\
x25519\_mlkem512 & 7,651 & 2,684 & 2,711 \\
p384\_mlkem768 & 30 & 149 & 265 \\
x448\_mlkem768 & 1,231 & 599 & 598 \\
x25519\_bikel1 & 23 & 390 & 28 \\
X25519MLKEM768 & 4,908 & 2,560 & 2,819 \\
X448MLKEM1024 & 1,092 & 593 & 995 \\
SecP256r1MLKEM768 & 4,832 & 2,488 & 2,668 \\
SecP384r1MLKEM1024 & 282 & 149 & 289 \\
\midrule
\textbf{Digital Signature Algorithm} & \textbf{Keygen/s} & \textbf{Sign/s} & \textbf{Verify/s} \\
\midrule
p256\_mldsa44 & 33 & 969 & 1,912 \\
rsa3072\_mldsa44 & 1 & 61 & 1,980 \\
p256\_falcon512 & 20 & 1,352 & 2,554 \\
rsa3072\_falcon512 & 1 & 63 & 2,674 \\
p256\_falconpadded512 & 20 & 1,351 & 2,557 \\
rsa3072\_falconpadded512 & 1 & 63 & 2,680 \\
p256\_OV\_Is & 21 & 2,077 & 1,699 \\
p256\_OV\_Ip & 27 & 3,361 & 1,748 \\
p256\_OV\_Is\_pkc & 22 & 2,012 & 317 \\
p256\_OV\_Ip\_pkc & 27 & 3,121 & 468 \\
\bottomrule
\end{tabular}}
\end{minipage}
\end{table*}

%##########################################################################################
\section{Discussion}
In the computational performance evaluations, lattice-based ML-KEM and ML-DSA variations consistently ranked among the top performers across both x86 and ARM systems. These algorithms often demonstrate low average operation times and reduced memory consumption. Furthermore, several candidates from the NIST additional signatures competition achieved high rankings compared to those that have been standardised when using lower security levels. For ARM, it is seen that the number of TLS handshakes decreases from 5,530 for ML-DSA-44/ML-KEM-512 to 3,672 for ML-DSA-87/ML-KEM-1024. This is a reduction of around 34\% in performance throughput, while the decrease on the x86 processor is comparatively smaller. 

A similar pattern was observed in the TLS performance testing, where PQC and Hybrid-PQC variations of ML-KEM and ML-DSA often provided higher performance, even when compared to classical schemes. Interestingly, lower security variations of CROSS and UOV also performed well on x86 systems, showing higher total connections than various classical schemes when using ML-KEM-512 as its KEM algorithm. In contrast, ARM systems did not exhibit this trend, with these signature schemes displaying reduced networking performance relative to ML-DSA, even at higher security levels.

These observations indicate that the standardised lattice-based PQC schemes provide reliable and consistent performance across system architectures. However, higher security levels exhibit reduced efficiency on ARM systems, which warrants further investigation if these schemes are to be deployed in security-critical environments. Furthermore, the promising results of several Additional Signature schemes suggest that continued research into their performance characteristics is necessary to gain a clearer understanding of their deployment requirements.

Overall, the collected evaluation results effectively demonstrate the framework's ability to capture and process PQC computational and networking performance data. By automating environment setup, test execution, and data parsing, the framework reduces manual intervention and improves performance testing execution. Its compatibility for both x86 and ARM evaluation enables performance comparisons across a range of hardware, which is particularly valuable for assessing PQC adoption in applications such as IoT.
%##########################################################################################
\section{Conclusion}
This paper introduced the PQC-LEO framework, an automated benchmarking suite for evaluating the computational and networking performance of PQC algorithms. By providing tools for environment setup, performance testing, and result parsing, the framework lowers the entry barrier for researchers and developers who wish to evaluate the challenges associated with PQC adoption. Its cross-platform compatibility with x86 and ARM architectures enables comprehensive evaluation of PQC performance in a range of deployment contexts. A proof-of-concept evaluation was conducted on both supported system architectures to demonstrate the framework's functionality and use case. Performance data was collected for CPU and memory metrics, as well as TLS 1.3 performance for PQC, Hybrid-PQC, and classical algorithms across physically networked devices.

Future development will expand functionality and device support. Planned enhancements include broader OS compatibility, integration of additional PQC libraries such as pqm4 for embedded platforms, automated energy consumption analysis, and extended TLS testing to capture metrics such as handshake sizes and simulated congestion. These enhancements will enable PQC-LEO to support more comprehensive assessments of PQC performance.

% Future development will expand functionality and device support. Planned enhancements include broader OS compatibility, integration of additional PQC libraries such as pqm4 for embedded platforms, automated energy analysis, and extended TLS testing to capture metrics like handshake sizes and simulated congestion. These improvements will enable more comprehensive assessments of PQC performance and overhead.

% Future development of the framework will focus on expanding its functionality and device support. Planned enhancements include broader operating system compatibility beyond Debian-based systems and integration of additional PQC libraries, such as pqm4, to support evaluation on embedded and microcontroller platforms. Further improvements will also include automated measurement and analysis of energy consumption, enabling more comprehensive assessments of PQC performance in resource-constrained environments. Additionally, TLS performance testing will be extended to capture a wider range of metrics, including handshake packet sizes and support for simulating network congestion. These enhancements will provide a comprehensive understanding of the communication overhead introduced by PQC algorithms and improve the framework's ability to support decisions regarding PQC adoption.

%----------------------- ---------------------------------------------------------------
% References
\bibliographystyle{IEEEtran}
\bibliography{main}

\end{document}